# "It's Viral!" - A study of the behaviors, practices, and motivations of TikTok Social Activists

Daniel Le Compte

Human-Computer Interaction Institute, Carnegie Mellon University, daniellecompte@cmu.edu

Daniel Klug

Institute for Software Research, Carnegie Mellon University, dklug@cmu.edu

Social media platforms such as Facebook and Twitter are used for social activism purposes, and TikTok is no different. We conducted 9 qualitative semi-structured interviews with social activists who recently posted their videos on TikTok to understand. This study presents an initial look into why TikTok is used by social activists, and what processes they use to carry out this work. The interviews revealed the following main patterns: (1) content creation practices are typical and expected, (2) motivation and inspiration for posting social activist content comes from a wide range of personal sources, (3) social activism has communities on TikTok that provides encouragement and discussion, (4) engagement and interaction with other activists and viewers is a crucial part of content creation as well as social activism, and (5) the main driving force behind picking TikTok over other platforms is the ability to spread messages much farther with less effort. These findings provide insight into the unique factors that TikTok brings for social activists and corroborates previous findings in understanding how social activists use social media for their purposes.

## 1 INTRODUCTION

Social media platforms serve as "networked publics", where there is a blend of technology, people, and practice, forming a unique space to allow people to gather for social, cultural, and civic purposes that have special affordances that other "publics" do not offer (boyd, 2010). Specifically, a unique purpose of networked publics is social activism, seen on many social media platforms such as Facebook, Twitter, and now TikTok, increasing drastically over the past 15 years (Rotman et al., 2011). TikTok in particular has emerged as a unique social media platform that centers user-generated content as its primary entertainment, discussion piece, and form of interaction (Kaye et al., 2020; Shutsko, 2020). This interaction and discussion largely occurs through three main features that, in some ways, resemble forums or threads on other platforms: Duet, Stitch, and "Comment in video" (e.g., Duet example, Stitch example, Comment in video example). These forms of interaction are crucial for social activism purposes, such as education on new topics, prompting of collective action, and discourse on divisive issues.

## 1.1 Social Activism on Social media

While activism has always relied on social media in some form to get the message out and to encourage action, in recent years social activism has moved largely online on social media platforms as mainstream media formats have not suited activists efforts or properly represented their views well (Gameson & Wolfsfeld, 1993; Watkins, 2001).

More specifically, previous research has noted the various dynamics and affordances that different social media platforms have may differ or depend on which platform is being used, such as a "Reddit-like forum" serving as a central hub for communication and organizing for movements like Anti-Extradition Bill Movement in Hong Kong (Lee et al., 2021). This dynamic between different platforms can be seen in multiple examples, such as using email chains and Twitter to put together a cultural and literacy festival using decentralized networking (Choi et al., 2019), or the many cases of Twitter being used for conversation and dialogue on political topics (Barberá et al., 2015), coordinated uprisings in the "Arab Spring" (Starbird & Palen, 2012), immigrant activism (Harlow & Guo, 2014), organization and action calling through Twitter (Choi et al., 2020), neighborhood community engagement via Facebook (Mosconi et al., 2017), and Twitter being used as a place for discussion, entertainment and humor in these movements (Freelon et al., 2016). These examples showcase the many usages of social media for social activism, ranging from educational purposes, opinionated discussion and discourse, action calling and organizing movements, to humor, entertainment, and place-finding.

In short, it is clear that social media platforms are utilized for a numerous set of purposes in social activism, and that these purposes are fulfilled on several different platforms, such as Twitter, Facebook, Email networks, blogs and more. What is not fully understood, as of yet, is how TikTok plays a role in this system as a newer social media platform with unique characteristics.

## 1.2 Social Activism on TikTok

TikTok has been available worldwide since 2017 however, only in the past two years, or so, has the academic community started to study how TikTok is used to fulfill various social activist purposes. Initially, we can see cursory and exploratory studies evaluating aspects of action and organization, such as in the case of #TulsaFlop with activists working against the Trump2020 campaign (Bandy & Diakopoulos, 2020), or in the initial findings of how TikTok might be used by various groups for discussion, venting, and finding community (Subramanian, 2021). Additionally, there have been some studies into how political discourse takes shape on TikTok (Serrano et al., 2020). Lastly, another poignant example of the usage of Tiktok for social movements and awareness on topics is a TikTok creator using her makeup videos as a platform to talk about the treatment of Uighurs by China in late 2019 (Kuo, 2019). While these studies show the initial usage of TikTok for various social activist purposes, many only use TikTok videos and other reports as data, and take a surface-level look at the content itself, as opposed to looking into the motivations behind this type of content or how this content comes to fruition. Based on this, our research was guided by the following questions: Why and how do social activists in general use TikTok for social activism? Do they use typical practices that are found elsewhere on TikTok, including popular or trending features? Why do they use TikTok over other social media platforms? What role does social activism on TikTok play in the greater work of social activism they do?

## 2 METHOD: DATA COLLECTION AND ANALYSIS

We conducted qualitative semi-structured interviews with nine TikTok users between March and April 2021 . Participants had to have posted at least one video in the last month on some form of a social activist message. Participants were found by searching for popular activism hashtags on TikTok, such as #socialactivism, #stopasianhate, #blm, #activism, or #environmentalism. Participants were aged between 18 and 30 and as of May 2021 they had 30 to 250.5k followers, had posted 7 to 100+ TikToks, and had 185 to 8.6m likes on their TikTok videos. Interview questions were broken up into four sections: About participants' background as social activists, how they utilized TikTok, why they used TikTok for social activism, and how social activism on TikTok fits into their overarching process of activism.

The analysis of interviews followed an open coding process in which initial codes were generated based on a generative, build-up approach to coding, similar to other studies and methods found in HCI and CSCW research (Mcdonald et al., 2019). This allows themes to emerge without preexisting patterns and questions to determine the direction of the data, similar in structure to a Grounded Theory approach, slimmed down for application in this project (Smith et al., 1996). Codes generated from one interview would be utilized in the next, though new codes would be created in the event of new or unique perspectives and responses from participants found in the transcripts. The results of this submission/contribution are based on six out of the nine interviews with a plan to code the remaining three interviews. This process led to the creation of 135 of codes, applied to 248 excerpts of transcripts, with 9 of categories of codes created.

## 3 RESULTS

From the interviews, there are five main themes that describe how and why TikTok users post social activist content on TikTok: (1) content creation practices are typical and expected, (2) motivation and inspiration for posting social activist content comes from a wide range of personal sources, (3) social activism has communities on TikTok that provides encouragement and discussion, (4) engagement and interaction with other activists and viewers is a crucial part of content creation as well as social activism, and (5) the main driving force behind picking TikTok over other platforms is the ability to spread messages much farther with less effort.

### 3.1 Creation practices are typical and expected

Throughout all interviews, participants described creation practices that are typical of most TikTok users. Ranging from improv and "off the top-of-my-head" filming and creation, to more extensive practices that require preparation and practice, the creation practices of interviewees are wide ranging yet typical in most senses. One participant, in particular, exemplifies the casual, more improv user persona: "*Um, you pretty much just kind of it's, it's this idea, this concept that I have in my head and it's pretty much all improv I speak*" (P01). This exemplifies ideas of "Slacktivism" by Harlow & Guo (2014), where activism is gaining popularity in the world of simple, short, and low-effort methods to get a message out and to encourage participation. Additionally, this form of video creation follows a low-budget and minimal resources format: "(...) *and then the actual filming, I mean, it's a quite low budget. I mean, it's just kinda like my phone, um, either natural lighting or with a ring light that I've got*" (P05).

In addition to simple creation practices, almost all participants used some form of common editing techniques that are found on the TikTok platform, such as blending other content from outside of direct filming, dueting and stitching features, as well as using captions, images, and greenscreen effects to create their videos. Other participants, in addition, follow popular TikTok trends to get their messages across, such as filming practices and trends, trending audio, as well as trending hashtags.

Only for a few participants (P01, P03, P04) were lengthy or involved processes used to create their messages and content. One participant (P03) in particular used a near "academic" process to research, study, and collect information and evidence about a topic, creating a "Master Document" that noted references and various points of opinion and evidence, before creating a script for a multi-part series of videos, one of which had 15 parts. This action seems to be not the norm, but not out of the scope of what participants would call necessary or useful for their type of content.

### 3.2  Inspiration and motivation come from a wide range of sources

Across all participants, inspiration and motivation for content comes from multiple sources of a diverse range. For example, P03 describes the motivation to begin posting social activist content stemming from the events of April and May 2020: "*Um, back in, I think it was April or May of last year, I was feeling very powerless with everything that was going on in the world*". Another participant described their area of work being an area they felt that they could contribute to the conversation: "*Well, I can't talk that much about racism, (...) but I can talk about white saviorism because that's the industry that I live in*" (P02). Lastly, some participants described seeing a viewpoint or area of content on TikTok that they felt they needed to respond to as motivation for beginning to post social activist content to TikTok: "(...) *and that truly began because a mutual friend of mine started sharing Q Anon [content] all over his Instagram and he was great gaining a lot of traction and there was a lot of, um, anti mask protests popping up. And I was just like, I, I can't argue, I can't be bothered arguing in the comment section*" (P04). In many senses, the motivation for posting is deeply personal to participants, and is widely diverse and stems from many sources.

### 3.3  Social activism community on TikTok encourages discussion and engagement

In discussing why and how they started posting social activist content to TikTok, participants mentioned the notion of joining the social activist community on TikTok. One participant (P04) noted that the amount of content represented an opportunity to join "*Uh, but I saw so much social activism on there. I knew that there was an opportunity to do so there*". Whereas another noticed that there was a method and pattern of meeting other social activists on TikTok, "*And they've [TikTok] really been pushing the social activism on my page, which is really cool. Cause I'm able to meet all these other social activists out there who inspire me to make my material*" (P01). These comments point towards a connection between those who share similar values as an entrance point to the discussion as well as connections beyond the content and TikTok sphere.

### 3.4  Engagement and interaction with viewers is crucial

Across all participants was a desire to utilize many methods of interacting and engaging in discussion, discourse, and commentary on social activists topics and issues. For some participants, this became a core source of content and interaction with their audience members, such as one participant (P06) using suggestions from comments to create a "Abortion policy rating" system and video series for different countries. In their words of describing the reaction to this idea of interaction, "(...) *getting people to comment and having this sort of format where you it's very, it's like interactive and it's fun because it's like, 'Oh my gosh, she's gonna give, she's* *gonna rate a legislation. That's so weird'"* (P06). In this, they would take suggestions from their viewers to spark conversation and continue discussion on an important topic.

For other participants (P07, P08), the purpose of engagement and interaction with others on TikTok was largely a form of participating in social activism itself. One participant described their interactions as advocating for a particular belief or viewpoint: "*And I like to, um, advocate rather than discriminate. So if there's somebody who has a very rude comment, rather than backlashing them and calling them out and TikTok, I just explained to them like my side*" (P01).

This format of engagement and interaction over a central topic, such as the Black Lives Matter movement, occurs in a different way than on other social media platforms. On TikTok, participants in conversation typically create videos through Stitches, Duets, using Comments in videos, or new videos entirely, to participate, such as in the trending audio "I NEED YOU TO" by Tobe Nwigwe, where users will mouth the words to the audio while acting, and then will switch suddenly to comment "Arrest the killers of Breonna Taylor" (Nwigwe, 2021; Rosenblatt, 2020). This provides a unique avenue for discourse and fulfilling activists needs, such as education and prompting action.

### 3.5 The key reason for posting is the ability of TikTok to spread a creator's message far

When asking about the reason why participants use TikTok over other platforms for their social activist messages, one theme primarily emerged: TikTok as a platform allows for their content and messages to reach a much wider audience than what is possible with other platforms. Some participants noted that the usage of TikTok helps get their message out beyond their own "circle": "So I was able to focus what my following was for rather than, um, Facebook, where it's like just friends of friends or family people you meet in real life" (P04). A main limiting factor of other platforms, that participants noted, was the necessity for audience members to connect or follow a creator before they would be able to see the content, unless in the unlikely event that the content was "promoted" through ads, or went viral: "*Well, I feel like TikTok, at least when I started using it, I felt like it was a lot easier for-for my videos to get sent to random people. Which is true, like on Instagram, unless I pay for it*" (P05). In general, this feature and motivation for using TikTok was truly beneficial, as it also helped in the exchange of information and content between social activists and audience members alike: "(...) *cause it really helps other people get their material to me. And it helps me get my material to other people*" (P01). The motivation and reasoning for social activists to use TikTok connects to what may be found when we examine the early reasons for social activists to utilize social media in the first place. As noted by Watkins (2001), to enact change, social activists desire to get the message across that fits their needs to the people that need to see it the most and that will be swayed into action or agreement the most. For the participants in this study, TikTok fulfilled this role more than any other platform.

### 4 CONCLUSION AND FUTURE WORK

Overall, the findings of this study support previous research in attempting to understand how social activists use social media to carry out their efforts for social activism, such as education, discourse, and encouraging action. What this study brings new to the literature is a perspective on how and why social activists use TikTok in particular over other social media platforms, such as Twitter or Facebook. Of particular impact for designing social media platforms or for social activists and their communication strategies is the findings of engagement and interaction that the social activists find important and the motivation for using TikTok over other platforms. Future work includes analyzing the remaining interviews based on the existing codes, and assessing results for extended depth based on the full dataset. While this study provides a unique perspective on social activism on social media, there remains questions of how persistent or generalizable these findings may be, and how this recent shift to a new platform may remain stable or used over a longer period of time.